\documentclass[aip,jmp,showkeys,showpacs,preprint]{revtex4-1}

\DeclareMathAlphabet{\mathsf}{OT1}{cmss}{m}{it}

\usepackage{amsmath,amssymb}
\usepackage{graphicx}
\usepackage[mathscr]{eucal}
\newcommand{\NN}{\nonumber\\}
\newcommand{\eus}[1]{\mathscr{#1}}
\newcommand{\vct}[1]{{\mathchoice{\mbox{\boldmath$#1$}}{\mbox{\boldmath$#1$}}%
  {\mbox{\scriptsize\boldmath$#1$}}{\mbox{\scriptsize\boldmath$#1$}}}}
\newcommand{\sub}[1]{_{\mbox{\scriptsize#1}}}
\newcommand{\sur}[1]{^{\mbox{\scriptsize#1}}}
\newcommand{\U}[1]{{\def\O{\mbox{O}}\def\u{\mbox{u}}\mathcode`\u=\mu\mathcode`\O=\Omega\mathrm{#1}}}
\newcommand{\fracpd}[2]{\frac{\partial#1}{\partial#2}}
\newcommand{\diver}{\mathop{\mathrm{div}}}
\newcommand{\curl}{\mathop{\mathrm{curl}}}

\begin{document}

\title{Mathematical structure of unit systems}

\author{Masao Kitano}

\email{kitano@kuee.kyoyo-u.ac.jp}
\affiliation{Department of Electronic Science and Engineering,
Kyoto University, Katsura, Kyoto 615-8510, Japan}
\date{\today, Revised base on the referee report}

\begin{abstract}
We investigate the mathematical structure of unit systems and 
the relations between them.
Looking over the entire set of unit systems,
we can find a mathematical structure that is called {\it preorder}
(or {\it quasi-order}).
For some pair of unit systems,
there exists a relation of preorder such that
one unit system is transferable to the other unit system.
The transfer (or conversion) is possible only when all of 
the quantities distinguishable in the latter system 
are always distinguishable in the former system. 
By utilizing this structure, 
we can systematically compare the representations in different unit systems. 
Especially, the equivalence class of unit systems (EUS) plays an important role 
because the representations of physical quantities and equations are of 
the same form in unit systems belonging to an EUS\@. 
The dimension of quantities is uniquely defined in each EUS\@.
The EUS's form a {\it partially ordered} set.
Using these mathematical structures, unit systems and EUS's are systematically
classified and organized as a hierarchical tree.
\end{abstract}



\maketitle

\section{Introduction}
A unit system is not simply a collection of units but 
an organized structure that enables 
diverse physical quantities to be represented in a unified manner.
In order to define a unit system, 
we first select a few units (quantities), 
which are referred to as {\it base units}. 
Other units are expressed as products or quotients of the base units 
and are referred to as {\it derived units} \cite{si, gupta}.
It is a matter of choice how many base units are selected.
A unit system with $N$ base units is called an $N$-base unit system.

With regard to unit systems, there are
several naive questions: 
How should the base units be chosen;
How many base units should be chosen;
How can we convert systematically from one unit system to another;
What is the meaning of dimensions of quantities (length, time, and so on)
\cite{birge, birge2, sommerfeld, duff, hehl, guggenheim}.

In principle, the number and choice of base units are
somewhat arbitrary.
This is the reason why there has been proposed so many unit systems and
standardization is strongly needed to avoid tangling of them.
Many modern articles on unit systems are focused on the
unification of unit systems or on the International System of Units (SI)
\cite{si, gupta}.

In the present paper, from more general point of view,
we investigate the mathematical structure of
unit systems and clarify the building principles and
the relationship between them.

We will show that a binary relation exists between
unit systems.
For certain pairs of unit systems, 
one of them can be derived 
or is {\it transferable} from the other.
The transferability relation satisfies the mathematical
axioms of {\it preorder} (or {\it quasi-order})
\cite{roman,bourbaki}.
For a given pair of unit systems, the possibilities are:
1) one of the unit systems is transferable to the other unit system,
2) both unit systems are transferable to each other (equivalent), or 
3) neither unit system is transferable to the other unit system (incomparable). 
It will be shown that the sorting of unit systems according to this preorder is
much more significant than the simple sorting by
the number of base units.

Especially, the equivalent case is important
because with this relation of equivalence we can
classify unit systems into equivalence classes.
We call such a class as
an {\it equivalence class of unit systems} (EUS). 
We will also find that the set of EUS's is a {\it partially ordered} set.
We can draw a hierarchical tree of unit systems and
EUS's by using the preorder and partial order structures. 
The structure of orders and equivalence greatly helps us to
sort out many existing and proposed unit systems.
There have been few such general and systematic study on
unit systems.

We will find that representations of 
physical quantities and equations have the same form
in unit systems belonging to an EUS\@.
Therefore, the EUS is a proper arena for quantity calculus,
which is a very important tool in science and engineering.
Quantity calculus is closely connected with
dimensional analysis \cite{buckingham, bridgman, bluman, porter}.

Originally the dimensions of units or quantities are introduced 
in order to cope with the situation where various units were
used for length, mass, and time \cite{maxwell1,porter}.
For example, the speed of light can be expressed 
in different units as,
$c_0 = 3\times10^8\,\U{m/s}=6.71\times10^8\,\U{\text{miles}/\text{hour}} = \cdots$.
We note that the unit for velocity is always expressed as
a unit of length divided by a unit of time.
We can write, 
independently of units,
the dimension of $c_0$ as
$\mathsf{L}^1 \mathsf{M}^0 \mathsf{T}^{-1}$,
with
the dimensions for 
length $\mathsf{L}$,
mass $\mathsf{M}$, 
and time $\mathsf{T}$.
In electromagnetism, however, the situation becomes complicated.
For example, in the meter-kilogram-second-ampere (MKSA) system,
the dimension for electric charge is
$\mathsf{L}^0\mathsf{M}^0\mathsf{T}^1\mathsf{I}^1$,
where $\mathsf I$ is the dimension for electric current,
while in the centimeter-gram-second (CGS) Gaussian system,
it is
$\mathsf{L}^{3/2}\mathsf{M}^{1/2}\mathsf{T}^{-1}$.

Thus the notion of dimensions is in a somewhat ambiguous situation.
It has been introduced to be independent of units 
but in fact depends on unit systems.
This situation has been noticed in many articles\cite{guggenheim,bluman,porter},
but no satisfactory explanation has been given.
In this paper, we will show that the dimension is uniquely defined in each EUS
but it is dependent on EUS's.
The above contradiction can be solved if we understand that
the MKSA and CGS Gaussian systems 
belong to different EUS's, while
the mechanical MKS and CGS unit systems
(and other mechanical unit systems) belong to a same EUS\@.

In this paper, we mainly use electromagnetic unit systems as examples,
because a rich variety of unit systems helps us to fully understand the 
present theory.
It will be easy to apply the theory to other fields.

We only deal with scalar quantities.
Generalization to geometric quantities such as vectors, tensors,
and differential forms can be made \cite{post,schouten}.
These multi-component quantities can be constructed from dimensioned scalars
as in mathematics these are derived from real numbers.

\section{Basics of unit systems}

\subsection{Ensembles of quantities}

We consider an ensemble $\varOmega$ that
contains all of the {\it physical quantities} under consideration.
At this stage, we make minimum assumptions on the physical
quantities in order to clarify the mathematical structures of unit systems.
We assume that for an arbitrary quantity $Q\in\varOmega$, the quantity
$cQ$, which is scaled by a real number $c\in\mathbb R$,
is also contained in $\varOmega$.
For such a scaled pair, $Q_1$ and $Q_2=cQ_1$, we define the sum as
$Q_1+Q_2=(1+c)Q_1$ and call that $Q_1$ and $Q_2$ are addible in $\varOmega$.
Negative quantities and subtraction can be considered with $c < 0$.
A sum is not defined for unscaled pairs.

We also assume that for any pair of nonzero quantities $Q, P\in\varOmega$, and
for any pair of rational numbers $\alpha,\beta\in\mathbb Q$, 
the quantity $Q^\alpha P^\beta$ is contained in $\varOmega$.
In other words, a product, a quotient, or a (fractional) power of quantities 
are defined.

\subsection{Representation of quantities with base units}

We now examine the role of the unit system.
To define a unit system, $U$, 
we choose $N$ quantities
$u_i\in\varOmega$ $(i=1,2,\ldots,N)$
that are to be referenced in the measurement
of general quantities.
These quantities are customary referred to as base units.
The set of base units is represented by a vector, $\vct u=(u_1,u_2,\ldots,u_N)$.
A physical quantity $Q\in\varOmega$ is represented as
\begin{align}
Q_U = q_U \vct u^{\vct d}
,
\label{eq:1}
\end{align}
where
$q_U=\{Q\}_U\in\mathbb R$ represents the {\it numerical value}
and
$\vct u^{\vct d}:=\prod_{i=1}^N u_i^{d_i}=u_1^{d_1}u_2^{d_2}\cdots u_N^{d_N}=[Q]_U$ 
represents the unit.
We refer to 
$\vct d=(d_1,\ldots,d_N)\sur T\in\mathbb Q^N$ as the 
{\it dimensional exponents}
of $Q$ in the unit system $U$.
(The unfamiliar notation $\vct u^{\vct d}$ is
borrowed from the notation $\vct x\cdot\vct y=\sum_{i=1}^N x_i y_i$
for the vectorial inner product.)

For example, the magnetic flux quantum
$\varPhi_0(=\hbar/2e)$,
defined in terms of Planck's constant $\hbar$ and
the elementary charge $e$,
can be represented in
the MKSA system with $\vct u=(\U{m},\U{kg},\U{s},\U{A})$
as $\varPhi_{0U}=2.07\times10^{-15}\,\U{m\,kg\,s^{-2}A^{-1}}$,
where 
$\{\varPhi_0\}_U=2.07\times10^{-15}$,
$[\varPhi_0]_U=\U{m\,kg\,s^{-2}A^{-1}}$,
and $\vct d=(1,1,-2,-1)\sur T$.

Note that Eq.~(\ref{eq:1}) is just a {\it representation},
which is dependent on the unit system,
and $Q_U$ does not designate the quantity itself.
\cite{birge, bridgman}%

The representation 
$Q_U=q_U\vct u^{\vct d}=(q_U, \vct d)\in \mathbb R\times\mathbb Q^N$,
is derived from the corresponding quantity $Q\in\varOmega$.
The {\it mapping}  
$\mathcal U: (Q\in\varOmega)\mapsto
(Q_U\in\mathbb R\times\mathbb Q^N)$ 
must satisfy the following properties:

\begin{enumerate}
\item
Each base unit $u_i\in\varOmega$ is mapped as,
$\mathcal U(u_i)=1\times u_i^1$.
\item 
For any $Q\in\varOmega$ 
with $\mathcal U(Q)=q_U\vct u^{\vct d}$,
and any $c\in\mathbb R$,
\begin{align}
\mathcal U(cQ) = c\,\mathcal U(Q) = (cq_U) \vct u^{\vct d}.
\label{eq:2}
\end{align}

\item
For nonzero quantities $Q,P\in\varOmega$ with representations
$\mathcal U(Q) = q_U \vct u^{\vct d}$ and
$\mathcal U(P) = p_U \vct u^{\vct b}$, 
and for
$\alpha,\beta\in\mathbb Q$,
the quantity $Q^\alpha P^\beta$ is represented as
\begin{align}
\mathcal U(Q^\alpha P^\beta)=
\mathcal U(Q)^\alpha \mathcal U(P)^\beta
=
(q_U^\alpha p_U^\beta)\vct u^{\alpha\vct d+\beta\vct b}
.
\label{eq:3}
\end{align}
Here, we can consider $\vct u^{\alpha\vct d+\beta\vct b}$ as a
unit for measuring $Q^\alpha P^\beta$.
A unit system that conforms to this condition is said to be {\it coherent}.

\item
If $Q_1$ and $Q_2$ are addible in $\varOmega$,
then $Q_1$ and $Q_2$ have the same dimension $\vct d$,
and we have
$\mathcal U(Q_1+Q_2)=\mathcal U(Q_1) + \mathcal U(Q_2)
=(q_{1U}+q_{2U})\vct u^{\vct d}$.

\item
Even when $Q$ and $P$ are not addible in $\varOmega$,
they may have the same dimension $\vct d$.
In this case, we can
write 
$\mathcal U(Q)+\mathcal U(P)=(q_U+p_U)\vct u^{\vct d}$,
i.e., $Q$ and $P$ become addible in the unit system $U$.
The addibility is not universal, but unit-system dependent.
\end{enumerate}

The mapping $\mathcal U$ is assumed to be surjective,
namely, for any $q_U\in\mathbb R$ and $\vct d\in\mathbb Q^N$,
there corresponds a quantity $Q\in\varOmega$ that satisfies
$\mathcal U(Q)=q_U\vct u^{\vct d}$.

Thus, the unit system $U=(\vct u,\mathcal U)$ is characterized
with the set of base units $\vct u$ and 
the mapping $\mathcal U:\varOmega\rightarrow \mathbb R\times\mathbb Q^N$.
We denote the number of base units as $N=\# U$.

\section{Preorder of unit systems}

\subsection{Unit-system dependent distinguishability of quantities}

If, in a unit system $U$, the presentations of two quantities $Q$ and $P$ 
coincide, i.e., $\mathcal U(Q)=\mathcal U(P)$,
then we write
$Q\stackrel{U}{=}P$.
More specifically, $Q\stackrel{U}{=}P$ indicates 
that $q_U=p_U$ and $\vct d=\vct b$,
for 
$\mathcal U(Q) = q_U \vct u^{\vct d}$ and
$\mathcal U(P) = p_U \vct u^{\vct b}$.
Clearly, $Q=P$ in $\varOmega$ implies that $Q\stackrel{U}{=}P$, 
although the converse is not necessarily true.
Generally, $Q\stackrel{U}{=}P$ does not imply $Q\stackrel{V}{=}P$
in another unit system $V$.
Thus the equality ``$\stackrel{U}{=}$'' is dependent on unit systems.


We should be careful not to write an equation such as 
$Q_U=Q_V$, even when $Q_U=\mathcal U(Q)$ and
$Q_V=\mathcal V(Q)$ are derived from the same quantity
$Q\in\varOmega$.
Consider two quantities $Q_1,Q_2\in \varOmega$ that satisfy
$Q_{1U}\neq Q_{2U}$ and
$Q_{1V} = Q_{2V}$.
If we write $Q_{1U}=Q_{1V}$ and $Q_{2U}=Q_{2V}$,
then we obtain a contradictory result: $Q_{1U}=Q_{2U}$.
A similar situation arises for the matrix representation of
vectors, i.e., we cannot write $(x_1, x_2)=(x_1',x_2')$,
even when $\vct x = x_1\vct e_1 + x_2\vct e_2 = x_1'\vct e_1' + x_2'\vct e_2'$.

The relation ``$\stackrel{U}{=}$'' is an equivalence relation 
\cite{roman}.
{\it Symmetry}, {\it reflexivity}, and {\it transitivity} hold,
i.e., 
(1) $Q\stackrel{U}{=}Q'$ implies $Q'\stackrel{U}{=}Q$,
(2) $Q\stackrel{U}{=}Q$, and
(3) $Q\stackrel{U}{=}Q'$ and $Q'\stackrel{U}{=}Q''$ imply
$Q'\stackrel{U}{=}Q''$, for all $Q$, $Q'$, and $Q''\in\varOmega$.

\subsection{Transferability of unit systems}

For a certain pair of unit systems, $U$ and $V$, if
$
(Q\stackrel{U}{=}P)
\Rightarrow
(Q\stackrel{V}{=}P)
$
holds for any pairs of quantities, $Q$, $P\in\varOmega$, we then denote
\begin{align}
U\succsim V
.
\label{eq:4}
\end{align}
This relation means that
the quantities that are considered to be equal in $U$ are always 
considered to be equal in $V$.
In other words, two quantities that are distinguishable in $V$ are
always distinguishable in $U$.
Then, we say that the unit system $U$ is {\it transferable to} the unit system $V$, 
or $V$ is transferable from $U$.

The relation ``$\succsim$'' satisfies
the axioms of {\it preorder} (or {\it quasi-order});
reflexivity and transitivity.
Namely, 
(1) $U\succsim U$,
and
(2) $U\succsim U'$ and $U'\succsim U''$ imply
$U\succsim U''$, for all $U$, $U'$, and $U''$.
Thus, the set of unit systems is a preordered set (poset)
\cite{roman, bourbaki}.
This is the key to understanding the global structure
of unit systems.

When both $U\succsim V$ and
$U\precsim V$ are satisfied, i.e., $U$ and $V$ are
bilaterally transferable,
we write $U\sim V$, and $U$ and $V$ are called to be {\it equivalent}.

There may be cases in which
neither $U\succsim V$ nor $U\precsim V$ are satisfied,
namely, $U$ and $V$ are transferable in neither direction.
In this case, we write $U\parallel V$, and say that $U$ and $V$ are {\it incomparable}.
Moreover, if
$U\succsim V$ and $V\not\succsim U$, then,
$U$ is strictly transferable to $V$ and we write $U\succ V$.
The relations are listed in Table \ref{tab:1}.

\begin{table}
    \centering
\begin{tabular}{c@{\quad}|@{\quad}c@{\quad\quad}c@{\quad}}
\rule[-0.7ex]{0mm}{3.5ex}
&
$U\precsim V$
&
$U\not\precsim V$
\\
\hline
\rule[-0.7ex]{0mm}{3.5ex}
$U\succsim V$
&
$U\sim V$ ($N=M$)
&
$U\succ V$ ($N>M$)
\\
\rule[-0.7ex]{0mm}{3.5ex}
$U\not\succsim V$
&
$U\prec V$ ($N<M$)
&
$U\parallel V$ ($N<=>M$)
\end{tabular}    
\caption{
Four possible relations between two unit systems $U$ and $V$:
(strictly) transferable to ($\succ$),
(strictly) transferable from ($\prec$),
equivalent ($\sim$), and incomparable ($\parallel$).
The relations between $N=\# U$ and $M=\# V$, which are the numbers of base units,
are also listed.
}
\label{tab:1}
\end{table}

\section{Conversion of unit systems}

\subsection{Mapping from one unit system to another}

We consider two unit systems,
$U=(\vct u,\mathcal U)$ and $V=(\vct v,\mathcal V)$
and assume that $U\succsim V$.
We will show that only in such cases
there exists a mapping $\mathcal T$ from
$\mathcal U(\varOmega)=\mathbb R\times\mathbb Q^N$ to 
$\mathcal V(\varOmega)=\mathbb R\times\mathbb Q^M$.

First, as shown in Fig.~1,
we choose an arbitrary representation $Q_U\in\mathbb R\times\mathbb Q^N$
in $U$.
There is a non-empty {\it preimage} (inverse image) 
$\mathcal U^{-1}(Q_U) \subset\varOmega$ because $\mathcal U$ is surjective.
$\mathcal U^{-1}(Q_U)$ does not mean an inverse mapping but
just designates a set containing all quantities 
that is mapped to $Q_U$ with $\mathcal U$.
The quantities in $\mathcal U^{-1}(Q_U)$ cannot be distinguished in $U$.
By choosing a quantity $Q$ in $\mathcal U^{-1}(Q_U)$ and
mapping it with $\mathcal V$, we obtain $Q_V$.
Its preimage $\mathcal V^{-1}(Q_V)$ consists of the quantities
that cannot be distinguished in $V$. 
From the assumption that $U\succsim V$,
$\mathcal V^{-1}(Q_V)$ should include $\mathcal U^{-1}(Q_U)$,
i.e., $\mathcal V^{-1}(Q_V)\supseteq\mathcal U^{-1}(Q_U)$.
Therefore, 
for a given $Q_U$, $Q_V$ is uniquely determined with
$\mathcal V(\mathcal U^{-1}(Q_U))=Q_V$.
Thus, we obtain a mapping (surjection) 
$\mathcal T: Q_U\in\mathcal U(\varOmega) \mapsto Q_V\in\mathcal V(\varOmega)$, or
$\mathcal T: U\rightarrow V$.

The following relations hold for $\mathcal T$:
for $c\in\mathbb R$, and $Q_U$ in $U$,
$\mathcal T(cQ_U)=c\mathcal T(Q_U)$;
for $\alpha,\beta\in\mathbb Q$, and $Q_U$, $P_U$ in $U$,
$\mathcal T(Q_U^\alpha P_U^\beta)=\mathcal T(Q_U)^\alpha\mathcal T(P_U)^\beta$;
for $Q_{1U}$, $Q_{2U}$, which are addible in $U$,
$\mathcal T(Q_{1U}+Q_{2U})=\mathcal T(Q_{1U})+\mathcal T(Q_{2U})$.

Note that if $U\succsim V$, then $N\geq M$,
where $N=\# U$ and $M=\# V$.
Therefore, for $U\sim V$, we have $N=M$, and the mapping $\mathcal T$ is reversible.
For $U\parallel V$, no mapping exits, and no definite relation between
$N$ and $M$ exists.
(See Table \ref{tab:1}.)

\begin{figure*}
\centering
\includegraphics[scale=1]{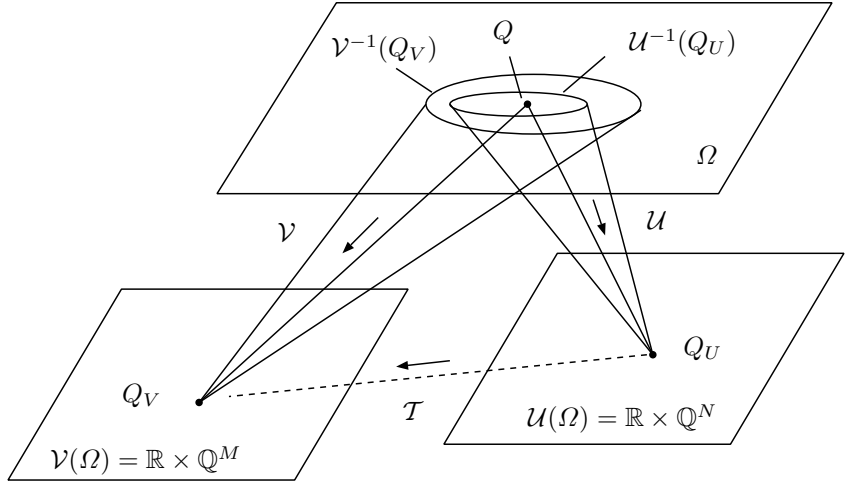}
\caption{
Ordering of unit systems $U$ and $V$.
A quantity $Q\in\varOmega$ is mapped to 
$\mathcal U(Q)$ and $\mathcal V(Q)$ for the representation in the
respective unit systems.
In the case of $U\succsim V$,
the preimage $\mathcal U^{-1}(Q_U)$
is always included in
$\mathcal V^{-1}(Q_V)$.
Only then we can naturally define the mapping
$\mathcal T: Q_U\mapsto Q_V
=\mathcal V(\mathcal U^{-1}(Q_U))
$.
}
\label{fig:1}
\end{figure*}

\subsection{Transfer matrix}
Let us consider two unit systems
$U=(\vct u,\mathcal U)$,
$\vct u=(u_1, u_2,\cdots, u_N)$
and
$V=(\vct v,\mathcal V)$,
$\vct v=(v_1, v_2,\cdots, u_M)$,
satisfying $U\succsim V$ and $N\geq M$.
A quantity $Q\in\varOmega$ is represented
in $U$ and $V$, respectively, as follows:
\begin{align}
\mathcal U(Q)=Q_U = 
q_U \vct u^{\vct d},
\quad
\mathcal V(Q)=Q_V = 
q_V \vct v^{\vct c}
,
\label{eq:5}
\end{align}
where
$q_U, q_V\in\mathbb R$,
$\vct d=(d_1, d_2, \ldots, d_N)\sur T\in \mathbb Q^N$, and 
$\vct c=(c_1, c_2, \ldots, c_M)\sur T\in \mathbb Q^M$.
As described in Sec.~IV A,
the relation between these representations can be
implemented as a mapping
$
Q_V = \mathcal T(Q_U)
$.

The explicit form of $\mathcal T$ can be obtained as follows.
Each base unit $u_i\in\varOmega$ $(i=1,\ldots,N)$ of $U$ can be
considered to be a representation in $U$ with
$\mathcal U(u_i)=u_{iU}=1\times u_i^1$, and can therefore be mapped by $\mathcal T$.
On the other hand, we have 
the representation of $u_i\in\varOmega$ in $V$ as
$\mathcal V(u_i)=k_i\vct v^{\vct t_i}$,
where
$k_i\in\mathbb R_+$ (positive real), $\vct t_i=(t_{1i},\ldots,t_{Mi})\sur T$,
$t_{ji}\in\mathbb Q$ $(j=1,\ldots,M)$.
From these expressions, we have
$
\mathcal T(u_{iU}) = 
k_i \vct v^{\vct t_i}
$.

Now we can map the representation
$\mathcal U(Q)=Q_U=q_U\vct u^{\vct d}$
of an arbitrary quantity $Q\in\varOmega$
as
\begin{align}
Q_V=\mathcal{T}(Q_U)
=\mathcal{T}(q_U\vct u^{\vct d})
=q_U\prod_{i=1}^N k_i^{d_i}\vct v^{\sum_{i=1}^N\vct t_i d_i}
= q_V\vct v^{\vct c},
\quad\text{with}\quad
q_V =q_U \vct k^{\vct d},\quad
\vct c = T\vct d
\label{eq:6}
,
\end{align}
where 
$
T
=
[\vct t_1,\vct t_2,\cdots,\vct t_N]
$
is an $M\times N$ matrix and $\vct k^{\vct d}:=\prod_{i=1}^N k_i^{d_i}=k_1^{d_1}\cdots k_N^{d_N}$. We refer to $T$ as a transfer matrix and assume that $\mathrm{rank}\, T=M$.

Thus, the mapping
$\mathcal T: Q_U\mapsto Q_V$ 
is characterized by
a vector $\vct k=(k_1, k_2, \ldots, k_N)\sur T\in\mathbb R_+^N$ and
a linear map $T\in L(\mathbb Q^N \rightarrow\mathbb Q^M)$.
Thus, we can write $\mathcal T=(\vct k, T)$.

Equation (\ref{eq:6}) indicates that, for $\vct d=0$, $Q_V=Q_U\,(=q_U)$ holds,
i.e.,
the dimensionless representations are conserved under the mapping.

\subsection{Composition of transformations}

The composition of transformations can easily be constructed.
Consider the mappings:
$\mathcal T=(\vct k, T)$ and
$\mathcal S=(\vct h, S)$,
with
$U\stackrel{\mathcal T}{\rightarrow}
V\stackrel{\mathcal S}{\rightarrow}
W$.
From $Q_V=\mathcal T(Q_U)=q_U\vct k^{\vct d}\vct v^{T\vct d}$ and
$\mathcal S(Q_V)=q_V\vct h^{\vct c}\vct w^{S\vct c}$,
we have the composite mapping:
\begin{align}
\mathcal S\mathcal T(Q_U)=q_U\vct k^{\vct d}\vct h^{(T\vct d)}\vct w^{S(T\vct d)}
=q_U(\vct k\vct h^T)^{\vct d}\vct w^{(ST)\vct d}.
\label{eq:7}
\end{align}
Here, we have used
\begin{align}
\vct h^{(T\vct d)}=\prod_{j=1}^M h_j^{\sum_{i=1}^N T_{ji}d_i}
=\prod_{i=1}^N \prod_{j=1}^M h_j^{T_{ji}d_i}=(\vct h^T)^{\vct d}
,
\label{eq:8}
\end{align}
with $(\vct h^T)_i=\prod_{j=1}^M h_j^{T_{ji}}$ $(i=1,\ldots,N)$,
and
$\vct k^{\vct d}\vct k'^{\vct d}=(\vct k\vct k')^{\vct d}$
with $\vct k\vct k'=(k_1 k'_1,\ldots,k_N k'_N)$.
From Eq.~(\ref{eq:7}), we have
the composition rule:
\begin{align}
\mathcal S\mathcal T =(\vct h^T\vct k, ST)
.
\label{eq:9}
\end{align}

We consider two invertible mappings $\mathcal T=(\vct k, T): U\rightarrow U'$,
$\mathcal S=(\vct h, S): U'\rightarrow U$.
The composite mapping
$\mathcal S\mathcal T=(\vct h^T\vct k, ST)$ becomes
the identity mapping
$\mathcal I=(\vct 1_N, I)$, if
$\vct h=\vct k^{-T^{-1}}$ and $S=T^{-1}$ are satisfied.
Here, we have introduced an $N$ dimensional vector 
$\vct 1_N=(1,1,\ldots,1)\sur T$.
In other words, the inversion of mapping $\mathcal T=(\vct k, T)$ is 
given as
\begin{align}
\mathcal T^{-1}=(\vct k^{-T^{-1}}, T^{-1}).
\label{eq:10}
\end{align}

\section{Quantities in equivalent unit systems}

\subsection{Equivalence class of unit systems}

In the case of $U\sim V$, the transformation $\mathcal T$ is
invertible and therefore the square ($N=M$) matrix $T$ is regular.
In this case, the two unit systems are basically the same because
there is a one-to-one correspondence between their representations,
and we can safely write $Q_U=Q_V$.

The relation ``$\sim$'' is an equivalence relation.
Therefore, according to this relation, we can classify $N$-base unit systems.
We refer to each class as an {\it equivalence class of unit systems} (EUS).
Unit systems that belong to different EUS's are incomparable.

If we do not transcend the border of an EUS,
then the representations $Q_U$ and $Q_V$ and the corresponding 
preimages $\mathcal U^{-1}(Q_U)=\mathcal V^{-1}(Q_V)$
in $\varOmega$ can be identified.
As almost unconsciously we are doing, we can write all of these representations as $Q$,
because there is no way to distinguish the members in the preimages within these unit systems.

For example, the MKS$\Omega$ (due to G. Giorgi) and MKSA systems are equivalent,
and we can write $1.2\,\U{O}=1.2\,\U{m^2 kg\, s^{-3} A^{-2}}\,(=R)$.
The CGS and MKS unit systems, both purely mechanical, are equivalent, and
we can write $1\,\U{erg}=10^{-7}\,\U{J}\,(=E)$.

\subsection{Relation among equivalence classes of unit systems --- partial order}

Let us consider a set of unit systems that is equivalent to a unit system $U$.
We write such an equivalence class of unit systems (EUS) as
$\eus{U}=\{U, U',\ldots\}$.
For any pair of the unit systems, there is an invertible mapping like
$\mathcal D = (\vct k, D):U\rightarrow U'$.
Then we have
\begin{align}
q_U \vct u^{\vct d}
= q_U \vct k^{\vct d}\vct k^{-\vct d}\vct u^{\vct d}    
= q_{U'}(\vct k^{-1}\vct u)^{\vct d}    
= q_{U'}(\vct k^{-1}\vct u)^{D^{-1}D\vct d}    
= q_{U'}\vct u'^{\vct d'}
,
\label{eq:11}
\end{align}
where $q_{U'}=q_U \vct k^{\vct d}$, 
$\vct d'=D\vct d$,
and
$\vct u'=(\vct k^{-1}\vct u)^{D^{-1}}$.
We note the these relations are invertible.
Using Eq.~(\ref{eq:10}), the last equation can be inverted with
$\vct k'=\vct k^{-D^{-1}}$.

Therefore, 
the representations in all the unit systems of an EUS can be identified as
\begin{align}
Q_{\eus U}=
q_U \vct u^{\vct d}=q_{U'}\vct u'^{\vct d'}=\cdots
.
\label{eq:12}
\end{align}
The collective expression $Q_{\eus U}$ is usually referred to as quantity,
which is believed to be independent of unit systems.
However, we now know that 
Eq.~(\ref{eq:12}) is valid only for the unit systems 
that belong to $\eus{U}$.
Therefore, we hereinafter refer to $Q_{\eus U}$ as {\it e-quantity} (quantity in an EUS).
In general, equations in physics represent relations
among e-quantities rather than mere quantities.
Therefore, such equations are valid only within an EUS\@.


If $Q$ and $P$ are addible in $U\in\eus{U}$, then they are addible in $U'\in\eus{U}$.
They are considered addible in $\eus U$.
The sum $Q_U+P_{U'}$, across the unit systems can be defined.

The binary relation $U\succsim V$ of preorder between unit systems can
be generalized to the binary relation $\eus U\succeq\eus V$ between
EUS's. For this relation, in addition to
reflexivity and transitivity,
{\it antisymmetry} is satisfied.
Namely, (1) $\eus U\succeq\eus U$, 
(2) $\eus U\succeq\eus U'$ and $\eus U'\succeq\eus U''$
imply $\eus U\succeq\eus U''$, and
(3) $\eus U\succeq\eus U'$ and $\eus U'\succeq\eus U$ imply
$\eus U=\eus U'$, for all $\eus U$, $\eus U'$, and $\eus U''$.
Such relations are referred to as {\it partial order} relations \cite{roman, bourbaki}.
Thus, the set of EUS's is a partially ordered set (poset).
This is also a very important view to understand dimensions and quantities
rigorously.

\begin{figure}
\centering
\includegraphics[scale=1]{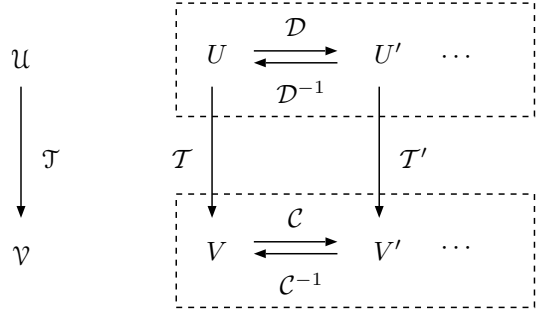}    
\caption{
Two equivalence classes of unit systems (EUS's) satisfying $\eus U\succ\eus V$, 
with the mapping $\eus T$. 
Each EUS contains equivalent unit systems as $\eus U=\{U, U',\ldots\}$,
$\eus V=\{V, V',\ldots\}$. 
There are invertible mappings between any pair of unit systems in an EUS, 
for example, $\mathcal D$ and $\mathcal D^{-1}$ between $U$ and $U'$. 
There is a (one-way) mapping from any unit system in $\eus U$ to any unit system in $\eus V$.
These mappings ($\mathcal T$ and $\mathcal T'$ in this example) are related as
$\mathcal T'=\mathcal C\mathcal T\mathcal D^{-1}$.
Note that there are also mappings between $U$ and $V'$ ($\mathcal D=\mathcal I$) or 
$U'$ and $V$ ($\mathcal C=\mathcal I$), which is not shown here.
}
\label{fig:2}
\end{figure}

As shown in Fig.~\ref{fig:2}, 
the mapping $\mathcal T:U\rightarrow V$ can be extended
to that between $\eus U$ and $\eus V$.
For $\eus U\succeq\eus V$, we have
$\eus T: Q_{\eus U}\in\eus U\mapsto Q_{\eus V}\in\eus{V}$.
The mapping $\mathcal T'$ between $U'\in\eus U$ and $V'\in\eus V$,
which is considered to be a representation of $\eus T$, can be
written as $\mathcal T' = \mathcal C\mathcal T\mathcal D^{-1}$
with $\mathcal D: U\rightarrow U'$, $\mathcal C: V\rightarrow V'$.

We denote $\eus U\parallel\eus V$, if $\eus U\not\succeq\eus V$ and
$\eus V\not\succeq\eus U$. 
We also denote $\eus U\succ\eus V$ if $\eus U\succeq\eus V$ and $\eus U\neq\eus V$.

\subsection{Dimension of e-quantities}
In this subsection, we explore the meaning of dimension
in detail using the framework of EUS\@.
Two quantities,
$Q_U=q_U[Q]_U=q_U \vct u^{\vct d}$ 
and 
$P_U=p_U[P]_U=p_U \vct u^{\vct c}$,
represented in $U$
are considered to have
a same dimension,
if they are represented by the
same unit;
$[Q]_U = [P]_U$,
or
$\vct d=\vct c$.
As long as we use only one unit system,
dimension is just a synonym of unit.
However, in another unit system $V \succ U$,
$Q$ and $P$ might be represented by different units.
Therefore, 
unlike commonly believed, 
dimension could be dependent on unit systems.

On the other hand, for an equivalent unit system
$U' \sim U$, $[Q]_{U'}=[P]_{U'}$ always follows
from $[Q]_U = [P]_U$, and vice versa.
Thus, we expect that the notion of dimension can be consistently extended
to every unit system belonging to the EUS.

Let us begin with a unit system $U\in\eus{U}$ with
base units $\vct u = (u_1, u_2,\ldots, u_N)$.
For each base unit $u_i$, we introduce a set
$\mathsf U_i = \{s_i u_i |  s_i \in \mathbb R_+\}$,
which is a collection of units different only in sizes.
Then, we make a unit system $U'$ with $\vct u' = (u'_1, u'_2,\ldots, u'_N)$,
$u'_i = s_i^{-1}u_i\in \mathsf U_i$ $(i=1,2,\ldots, N)$.
Each base unit of $U'$ is only different in size from the corresponding base unit of
$U$. 
The mapping is $\mathcal S = (\vct s, I): U\rightarrow U'$, 
$\vct s=(s_1, s_2, \ldots, s_N)\sur T$.
Therefore, $U'$ is
equivalent to $U$ and belongs to $\eus{U}$.
A quantity $Q$ can be expressed as
$Q_U=q_U\vct u^{\vct d}$ in $U$ and
$Q_{U'}=q_{U'}\vct u'^{\vct d}$ in $U'$, respectively,
and $\vct d$ is conserved under the scaling of units.

We can consider that two expressions 
$Q_U=q_U\vct u^{\vct d}$
and
$Q_{U'}=q_{U'}\vct u'^{\vct d}$
have a common dimension 
$\vct U^{\vct d}=\prod_{i=1}^N \mathsf U_i^{d_i}$,
where $\vct U = (\mathsf U_1,\mathsf U_2,\ldots,\mathsf U_N)$
is the dimensional basis.
For example,
$U=(\text{MKS})$, $\vct u=(\U{m}, \U{kg}, \U{s})$ and
$U'=(\text{CGS})$, $\vct u'=(\U{cm}, \U{g}, \U{s})$
share the dimension of the form
$\mathsf L^l\mathsf M^m\mathsf T^t$,
since $\U{cm}=10^{-2}\,\U{m}\in\mathsf L$,
$\U{g}=10^{-3}\,\U{kg}\in\mathsf M$, and
$\U{s}\in\mathsf T$.

More generally, for any $U'\in\eus{U}$, there exists an
invertible mapping $\mathcal D = (\vct k, D): U\rightarrow U'$ 
and 
$\vct U^{\vct d}=\vct U'^{\vct d'}$
holds, where
\begin{align}
\vct d' = D\vct d,\quad\vct U' = \vct U^{D^{-1}}
.
\label{eq:13}
\end{align}
As for dimensions, the scale factor $\vct k$ plays no role.
Under the transformation by any regular matrix $D$, 
the dimension $\vct U^{\vct d}$ is invariant, 
while $\vct U$ and
$\vct d$ transform in a reciprocal manner.
Dimension is conserved under invertible transformations of
unit systems. 
Dimension is invariant in the EUS, since any pair of unit systems
in an EUS can be related by an invertible transformation.
The dimension of an e-quantity in $\eus U$
can be represented collectively
as 
\begin{align}
[Q]_{\eus U}=\vct U^{\vct d}=\vct U'^{\vct d'}=\cdots
.
\label{eq:14}
\end{align}

We consider an EUS $\eus V$ that is not equivalent to $\eus U$.
The dimension of an e-quantity $Q_{\eus V}$ is expressed
as
$[Q]_{\eus V}= \vct V^{\vct c}$
with
$\vct V=(\mathsf V_1,\mathsf V_2,\ldots,\mathsf V_M)$.
If $\eus U\succ\eus V$, we have a non-invertible mapping
$\mathcal T=(\vct k, T):U\in\eus U\rightarrow V\in\eus V$, by which we can
convert the dimensions as
$\mathcal T(\mathsf U_i) = \vct V^{\vct t_i}$
and $\vct c = T\vct d$,
unidirectionally [See Eq.~(\ref{eq:6})].
Unlike the equivalent cases,
the former equations cannot be expressed as 
$\vct V = \vct U^{T^{-1}}$ because $T$ is not
invertible.
Therefore, we cannot equate
$\vct U^{\vct d}$ and
$\vct V^{\vct c}$.
In the case of $\eus U\parallel\eus V$, there even exist no such direct relations
between their dimensions.
Some examples on dimension in EUS will be given in Sec.~IX.G.

\subsection{Dimensional analysis and the Buckingham Pi-theorem}
The central result of dimensional analysis is
the Buckingham Pi-theorem\cite{buckingham, bridgman, bluman}.
It imposes restrictions on the form of equations that are physically sensible.
It also helps to extract non-dimensional parameters that characterizes
the problem under consideration.

We outline the proof of the theorem in the framework of equivalence
unit systems.
Let us consider a set of e-quantities 
$P_0, P_1, P_2, \ldots, P_L$ in $\eus{U}$
with $L > N := \#\eus{U}$.
We suppose they are related by a function $f$ as
\begin{align}
P_0 = f(P_1, P_2, \ldots, P_L)
.
\label{eq:15}
\end{align}
For simplicity, we omit subscripts $\eus{U}$ in e-quantities
in this subsection.
We assume that these e-quantities are arranged so that
$P_1, P_2,\ldots, P_N$ are dimensionally independent each other and
$P_0, P_{N+1},\ldots, P_L$ dependent on
$P_1, P_2,\ldots, P_N$.
We express the corresponding quantities in $U\in\eus{U}$ as 
$P_l = p_{lU}\vct u^{\vct d_l}$ 
$(l=0,1,\ldots,L)$.
In terms of dimensional exponents in $U$,
$\vct d_1,\vct d_2,\ldots,\vct d_N$ are linearly independents and
$\vct d_0, \vct d_{N+1},\ldots\vct d_L$ are linearly dependent on them, i.e.,
\begin{align}
\vct d_k = \sum_{n=1}^N \vct d_n T_{nk}
=D\vct t_k
\quad
(k= 0, N+1,\ldots, L)    
,
\label{eq:16}
\end{align}
where $D = [\vct d_1, \vct d_2,\ldots, \vct d_N]$
and $\vct t_k = (T_{1k}, T_{2k},\ldots, T_{Nk})\sur T$.
The coefficients $T_{nk}$ can be derived 
as $\vct t_k = D^{-1}\vct d_k$,
once the dimensional 
exponents $\vct d_l$ $(l=0,1,\ldots, L)$ are given.

With these we can make dimensionless e-quantities by normalization as
\begin{align}
\pi_k := 
\frac{P_k}{\vct P^{\vct t_k}}\in\mathbb R
\quad
(k= 0, N+1,\ldots, L)    
,
\label{eq:17}
\end{align}
where $\vct P=(P_1, P_2,\ldots, P_N)$.
Inserting into Eq.~(\ref{eq:15}), we have
\begin{align}
\pi_0 &= \vct P^{-\vct t_0} f(P_1, P_2, \ldots, P_N, \pi_{N+1}\vct P^{\vct t_{N+1}},
\ldots, \pi_L\vct P^{\vct t_L})    
\NN
&= F(P_1, P_2, \ldots, P_N, \pi_{N+1}, \ldots, \pi_L)
.
\label{eq:18}
\end{align}
By introducing appropriate (non-dimensional) function $F$, we can absorb
$P_n$'s in the normalization factors $\vct P^{\vct t_k}$ into the first $N$ arguments.

With $\vct P=(P_1, P_2, \ldots, P_N)$, we can form a basis of unit system
$\vct v=(v_1, v_2, \ldots, v_N)$ in $\eus{U}$ by rescaling of $v_n = P_n / p_{nV}$ with
arbitrary factors $p_{nV} \in\mathbb R_+$.
In this unit system $V$, the numerical parts of $P_n$ and $\pi_k$ are 
$\{P_n\}_V=p_{nV}$ $(n=1,2,\ldots, N)$ and $\{\pi_k\}_V = \pi_k$ $(k=0,N+1,\ldots,L)$.
Equation (\ref{eq:18}) should hold 
even if we replace the quantities with the corresponding
numerical factors as
\begin{align}
\pi_0 = F(p_{1V}, p_{2V}, \ldots, p_{NV}, \pi_{N+1}, \ldots, \pi_L)
.
\label{eq:19}
\end{align}
Since $p_{nV}$ can take any (positive) numerical values, 
the function $F$ should not depend on $p_{nV}$ and
Eq. (\ref{eq:19}) reduces to a relation among the dimensionless parameters:
\begin{align}
\pi_0 = G(\pi_{n+1},\pi_{n+2},\ldots,\pi_{L})
.
\label{eq:20}
\end{align}
Thus the dimensional consideration helps to simplify the forms of
physical equations.

\section{Standard form of transformation}

\subsection{Decomposition of the transfer matrix}

For the case where $U\succsim V$ holds but $U\precsim V$ does not,
i.e., the case of $U\succ V$,
the mapping $\mathcal T=(\vct k, T)$ is not invertible.
Then, $N>M$, and we set $L=N-M\geq 1$.

We can transform the matrix $T$ of rank $M$ into a standard form,
$J = \left[I_M |0\right]$ with an $M\times M$ matrix $C$ and 
an $N\times N$ matrix $D$, both of which are invertible,
as $T=C^{-1}JD$.
The $M\times N$ matrix $J$ is composed of the $M\times M$ unit submatrix $I_M$ and
the $M\times L$ zero submatrix \cite{maclane}.
Note that the matrix elements are all rational numbers.

We consider a series of mappings,
$U\stackrel{\mathcal D}{\rightarrow}U'
\stackrel{\mathcal J}{\rightarrow}V'
\stackrel{\mathcal C^{-1}}{\rightarrow}V$,
where $U'$ and $V'$ are $N$-base and $M$-base unit systems,
respectively.
The invertible mappings are defined as
$\mathcal D=(\vct 1_N, D)$ and
$\mathcal C=(\vct 1_M, C)$.
We also define a mapping $\mathcal J=(\vct k', J)$, in which 
$\vct k'=\vct k^{D^{-1}}$.
Using Eqs.~(\ref{eq:9}) and (\ref{eq:10}), we 
obtain the standard decomposition:
\begin{align}
\mathcal C^{-1}\mathcal J\mathcal D
=(\vct 1_M^{JD}\vct k'^{D}\vct 1_N, C^{-1}JD)
=(\vct k,T)=\mathcal T
.
\label{eq:21}
\end{align}

\subsection{Quantities transferred to unity}
The vectors $\vct d'_h=\vct e_{M+h}$ $(h=1,\ldots,L)$ 
belong to ${\rm ker}\, J$ and satisfy
$J\vct d'_h=0$,
where $\vct e_i\in\mathbb Q^N$ is the $i$-th unit vector.
Here, ${\rm ker}\, J$ represents the zero space (kernel) of $J$, i.e.,
the subspace spanned by the vectors $\vct d'$ with $J\vct d'=0$.
Then, the vectors 
$\vct d_h=D^{-1}\vct d'_h$ belong to ${\rm ker}\, T$ and
satisfy $T\vct d_h=0$.

Using $\vct d_h$, we can define the following representations in $U$:
\begin{align}
I_{hU}= \vct k^{-\vct d_h}\vct u^{\vct d_h}
,
\label{eq:22}
\end{align}
which is mapped to $V$ by $\mathcal T$ as
\begin{align}
I_{hV}=
\mathcal T(I_{hU})= \vct k^{-\vct d_h}\vct k^{\vct d_h}\vct v^{T\vct d_h}
= 1\vct v^0 = 1
.
\label{eq:23}
\end{align}
Thus, the representations $I_{hU}$ ($h=1,2,\ldots,L$)
are all considered to be unity in $V$.
The corresponding e-quantities $I_{h\eus U}$ 
are also considered to be unity in $\eus V$.

If we have two e-quantities $Q_{1\eus U}$ 
and $Q_{2\eus U}$, which are related in $\eus U$ as
$Q_{1\eus U} = I_{h\eus U}Q_{2\eus U}$, then $Q_{1\eus U}$ 
and $Q_{2\eus U}$ cannot be distinguished in $\eus V$
because of $I_{h\eus V}=1$.

More generally, the relation
$Q_{1\eus U} = I_{1\eus U}^{d_1}\cdots I_{L\eus U}^{d_L}Q_{2\eus U}$
with $(d_1,\ldots, d_L)\sur T\in\mathbb Q^L$,
reduces to $Q_{1\eus V}=Q_{2\eus V}$ in $\eus V$.
Therefore, the mapping $\eus T:\eus U\rightarrow\eus V$ is characterized
by the preimage
$\eus T^{-1}(1)=\{I_{1\eus U}^{d_1}\cdots
I_{L\eus U}^{d_L}\,|\,d_1,\ldots, d_L\in\mathbb Q\}$.

Now we know the implications of a shorthand method, in which some quantities
are considered to be unity, e.g., ``we set $c_0=1$.''

\section{Comparison of unit systems with normalized quantities}

\subsection{Normalized quantities}
As discussed earlier, we cannot directly equate the representations
in non-equivalent unit systems, even if each representation corresponds to the same quantity.
In order to overcome this inconvenience, we introduce the notion of {\it normalization}.
We assume that $\eus U\succ \eus V$ and show that
$\eus V$ can be embedded
into $\eus U$ by appropriately normalizing each e-quantity.

We use the standard decomposition (\ref{eq:21}).
In $U'\in\eus U$, $I_{h\eus U}$ is simply represented as 
$I_{hU'}=\vct k^{-\vct d_h}\vct u'^{\vct d'_h}
=(k_{M+h}')^{-1}u'_{M+h}$.
We have used Eq.~(\ref{eq:22}) and $\vct k'^{\vct d'_h}=\vct k^{\vct d_h}$.
For any $Q_{\eus U}$, which is represented as
$Q_{U'}=q_U \vct u'^{\vct d'}$
in $U'$, 
we introduce
a representation
$N_{U'}(Q_{\eus U})=I_{1U'}^{-d'_{M+1}}\cdots I_{LU'}^{-d'_{M+L}}$,
which satisfies $N_{V}(Q_{\eus U})=1$ and 
cancels the higher portion $(d'_{M+1},\ldots,d'_{M+L})$ of dimensional
exponent of $Q_{U'}$.
Then, we define a normalized representation of $Q_U$ in $U'$:
\begin{align}
\tilde Q_{U'}
=N_{U'}(Q_{\eus U})Q_U
=q_U(k_{M+1}')^{d_{M+1}'}\cdots(k_{M+L}')^{d_{M+L}'}
u_1'^{d'_1}\cdots u_M'^{d'_M}
.
\label{eq:24}
\end{align}
The normalized e-quantity $\tilde Q_{\eus U}=N(Q_{\eus U})Q_{\eus U}$ can be
represented only by the subset of base units:
$\tilde{\vct u}'=(u_1',\ldots,u_M')\subset\vct u'$.
Owing to $\vct v'=J\vct u'$, 
$\tilde{\vct u}$ is faithfully mapped to $\vct v'$:
$v_i'=\tilde u_i'$ $(i=1,2,\ldots,M)$.
This means that there is a one-to-one correspondence between
$\tilde{Q}_{U'}$ and $Q_{V'}$, or
between 
$\tilde{Q}_{\eus U}$ and $Q_{\eus V}$.
The normalization 
$Q_{\eus U}\mapsto \tilde Q_{\eus U}=N_{\eus U}(Q_{\eus U})Q_{\eus U}$ is found to be 
equivalent to the mapping $\eus T: Q_{\eus U}\mapsto Q_{\eus V}$.
Note that $\eus T(N_{\eus U}(Q_{\eus U}))=1$ for any $Q_{\eus U}$.

For $Q_{1\eus U}$ and $Q_{2\eus U}$, 
we can define the normalized e-quantities as 
$\tilde{Q}_{1\eus U}=N_{\eus U}(Q_{1\eus U})Q_{1\eus U}$ and
$\tilde{Q}_{2\eus U}=N_{\eus U}(Q_{2\eus U})Q_{2\eus U}$.
It is possible to render a situation
in which 
$\tilde{Q}_{1\eus U}=\tilde{Q}_{2\eus U}$
and
$Q_{1\eus U}\neq Q_{2\eus U}$ in $\eus U$.
Thanks to the normalization factors $N_{\eus U}(Q_{\eus U})$, 
we can keep track of the difference.

For the situation in which we need to clarify the unit system $V$
to which we move, we write $\tilde Q_U^V=N_U^V(Q)Q_U$,
and the same for the EUS,
$\tilde Q_{\eus U}^{\eus V}=N_{\eus U}^{\eus V}(Q)Q_{\eus U}$.

\subsection{Comparison of incomparable unit systems}

We now consider the situation in which we have to compare
unit systems $U$ and $V$, which are incomparable, i.e.,
$U\parallel V$.
These unit systems cannot be compared directly because,
in $U$ and $V$, the quantities are classified with different principles.
The normalization method only works for $U\succsim V$ or
$U\precsim V$.
Fortunately, we can handle this situation by finding a unit system
$W$ that is transferable to both $U$ and $V$, i.e., $W\succsim U$ and $W\succsim V$.
Then, we can normalize quantities as
$\tilde{Q}_W^U=N_W^U(Q) Q_W$ and
$\tilde{Q}_W^V=N_W^V(Q) Q_W$.

Thus, the representation in $U$ and $V$ can be
embedded into $W$ and can be considered as representations in $W$.

\section{Practical Unit Systems}

Historically, a number of types of unit systems have been proposed
and adopted but, at present, only a few of them are used \cite{gupta}.
This is partly because the use of the International System of Units
(SI) \cite{si}, which is an extended version of the 
MKSA system,
is strongly recommended and has gained popularity.
Even though a systematic study of unit systems may no longer appear to be necessary, 
we sometimes need to read articles and books
that are based on old unit systems and to convert
quantities from one unit system to another.
On such occasions, although conversion tables can be used,
there is no reliable way to confirm the correctness of the conversion.
Therefore, we need to have a rigorous theory of unit systems
so that we can confirm the accuracy of conversion tables in textbooks.
We can also logically assess presently used unit systems and compare them 
to systems that may be developed in the future.

In the this article, we deal with several electromagnetic
unit systems as examples.
The MKSA system, or the electromagnetic subset of the SI,
is a four-base unit system.
Hereafter, we use the terms MKSA and SI interchangeably.
The CGS emu (electromagnetic unit) system
and the CGS eus (electrostatic unit) system are both
three-base unit systems \cite{maxwell2}.
Normally, people use the non-rationalized versions of these systems in order to
simplify (or to remove the factor $4\pi$ from) the Coulomb and Biot-Savard laws.
However, the present consensus is that the {\it rationalized} system,
in which the factor $4\pi$ is moved to the field solutions for
point sources, is more reasonable.
Therefore, in the present article, in order to simplify the argument,
we use only {\it rationalized} systems
and denote these systems as rCGS-emu (emu for short) and rCGS-esu (esu).

The CGS Gaussian system is
a mixture of the CGS-emu and CGS-esu 
systems \cite{birge,jackson,kitano}.
We deal with only its rationalized version, which is referred to as
the Heaviside-Lorentz (HL) system \cite{sommerfeld}.
Moreover, we have to introduce a variant \cite{jackson}, 
which is modified to correct a defect of the HL system
as explained later. 
We hereafter refer to this version as 
the {\it modified} Heaviside-Lorentz system (mHL).

\subsection{Examples of the use of normalized quantities}

As emphasized repeatedly, we should be very careful not to
equate representations in different unit systems, such as 
$Q_U=Q_V$.
In the following, we further explore this point because,
although subtle, this is an important consideration.
The unit of electric current in the MKSA (or SI),
$I\sub{SI}=1\,\U{A}$, is represented in the rCGS-emu as
$I\sub{emu}=\sqrt{4\pi}\times 10^{-1}\,\U{\sqrt{dyn}}$ \cite{jackson}.
Even then, we should not write $I\sub{SI}=I\sub{emu}$, because
$1\,\U{A}=\sqrt{4\pi}\times 10^{-1}\,\U{\sqrt{dyn}}$ is dimensionally
inconsistent.
From the viewpoint of rCGS-emu, the left-hand side contains an undefined unit, ``A''.
From the viewpoint of MKSA, the equation
reduces to $1\,\U{A}=\sqrt{40\pi}\times 10^{-3}\,\U{\sqrt{N}}$, which is incorrect.

Let us consider this problem in more detail.
Since $I\sub{SI}/\U{A}=1$ and $I\sub{emu}/\U{\sqrt{dyn}}=\sqrt{4\pi}/10$,
we obtain the following dimensionless relation: 
\begin{align}
\frac{I\sub{emu}}{\U{\sqrt{dyn}}}=\frac{\sqrt{4\pi}}{10}\frac{I\sub{SI}}{\U{A}}
,
\label{eq:25}
\end{align}
which is valid for current of any amplitude.
This is the best we can do for representations of different unit systems.
We cannot multiply both sides by $\U{\sqrt{dyn}}$ or by $\U{A}$
in order to simplify the equations.
In the former case, we have a mixture of units on the right-hand side,
and in the latter case, we have a mixture of units on the left-hand side.
In order to proceed, we can use the normalization and have a relation
in the MKSA, 
\newcommand{\sstrut}{\rule{0mm}{1.3ex}}
\begin{align}
\tilde{I}\sur{emu}\sub{SI}=\sqrt{\sstrut\mu\sub{$0$,SI}} I\sub{SI}
,
\label{eq:26}
\end{align}
which corresponds to Eq.~(\ref{eq:25}).
Using
$\tilde{I}\sur{emu}\sub{SI}$ instead of $I\sub{emu}$,
we can legitimately multiply 
both sides by $\sqrt{\U{dyn}}=\sqrt{10^{-5}\,\U{N}}$.
Note that $\mu\sub{$0$,SI}=4\pi\times10^{-7}\,\U{N/A^2}$.

As another example, we consider a magnetic field strength $H$ and
a magnetic flux density $B$, each of which are represented in the MKSA and
rCGSemu.
If the relation 
\begin{align}
B\sub{SI}=\mu\sub{$0$,SI}H\sub{SI}
,
\label{eq:27}
\end{align}
is satisfied in the MKSA, then, in the rCGSemu, 
\begin{align}
B\sub{emu}=H\sub{emu}
,
\label{eq:28}
\end{align}
holds.
If we mistakenly write $B\sub{SI}=B\sub{emu}$ and
$H\sub{SI}=H\sub{emu}$, then we have
a contradictory relation $\mu\sub{$0$,SI}=1$.

Using the normalization
$\tilde{B}\sur{emu}\sub{SI}:=(1/\sqrt{\sstrut\mu\sub{$0$,SI}})B\sub{SI}$,
$\tilde{H}\sur{emu}\sub{SI}:=\sqrt{\sstrut\mu\sub{$0$,SI}}H\sub{SI}$,
we have
$\tilde{B}\sur{emu}\sub{SI}=\tilde{H}\sur{emu}\sub{SI}$,
which corresponds to Eq.~(\ref{eq:28}).
Similarly, for the rCGS-esu,
$\tilde{B}\sur{esu}\sub{SI}:=\sqrt{\sstrut\varepsilon\sub{$0$,SI}}B\sub{SI}$,
$\tilde{H}\sur{esu}\sub{SI}:=(1/\sqrt{\sstrut\varepsilon\sub{$0$,SI}})H\sub{SI}$,
we have 
$\tilde{B}\sur{esu}\sub{SI}=(1/c_{0,SI}^2)\tilde{H}\sur{esu}\sub{SI}$.

The next example is to compare the representation of a charge in the esu and emu.
We have
$\tilde q\sur{esu}\sub{SI}=q\sub{SI}/\sqrt{\varepsilon\sub{0SI}}$ and
$\tilde q\sur{emu}\sub{SI}=\sqrt{\mu\sub{0SI}}q\sub{SI}$,
the units of which are $\U{\sqrt{N}\, m}$ and $\U{\sqrt{N}\, s}$, respectively. 
From these we obtain the notable Weber-Kohlrausch relation \cite{maxwell2} as follows:
$\tilde q\sur{esu}\sub{SI}/\tilde q\sur{emu}\sub{SI}=c\sub{0SI}$.
Thus the MKSA system serves as a framework for comparing
the rCGS-emu and rCGS-esu systems.

\begin{figure*}
\centering
\includegraphics[scale=1]{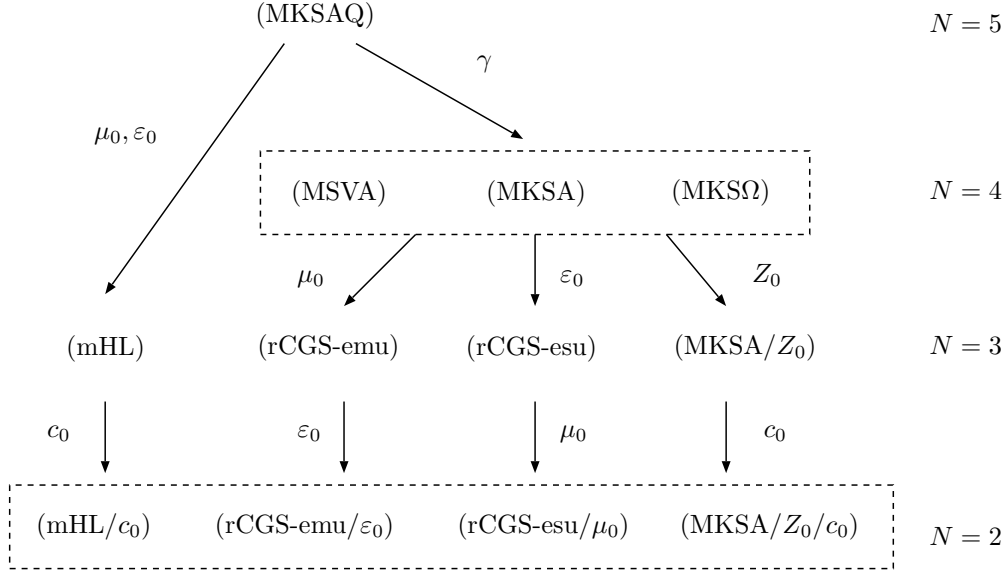}  
\caption{
A hierarchical tree of unit systems. 
Here, $N$ is the number of base units.
Arrows indicate transferability ``$\succ$'', and the
associated quantity is considered to be unity on the transfer.
Dashed boxes represent EUS's, and 
the four- and two-base unit systems listed are equivalent within each group,
whereas the three-base unit systems are all incomparable.
} 
\label{fig:3}
\end{figure*}

\section{Relations between real unit systems}

In the following, we compare several unit systems,
some of which are practically used systems.

\subsection{Example 1}
Let us start with a toy model.
We consider a set $\varOmega$, which includes quantities for 
voltage and current.
In a unit system $U$, we use the ampere and the volt as
the base units,
$\vct u=(\U{A},\U{V})$,
and in the other system $V$, the watt and the ohm,
$\vct v=(\U{W},\U{O})$.
We have
$\mathcal T(\U{A})=1\,\U{W^{1/2}O^{-1/2}}$,
$\mathcal T(\U{V})=1\,\U{W^{1/2}O^{1/2}}$, 
and find $\mathcal T=(\vct k, T): U\rightarrow V$ as
\begin{align}
\vct k = (1,1),
\quad
T=
\begin{bmatrix}
1/2 & 1/2
\\
-1/2 & 1/2    
\end{bmatrix}
.
\label{eq:29}
\end{align}
We see that
$\text{ker}\, T=\{0\}$, i.e., 
$\mathcal T$ is invertible, and
$U\sim V$.

\subsection{Example 2}
As another simple example,
we consider a set $\varOmega$, which includes quantities for
time and length.
In $U$, we adopt the base units
$\vct u=(\U{m},\U{s})$ and in $V$ we use
$\vct v=(\U{m})$.
In the latter, time is measured in terms of length with the
help of the speed of light $c_0$.
We have
$\mathcal T(\U{m})=1\,\U{m}$,
$\mathcal T(\U{s})=\{c_0\}_U\,\U{m}$, 
where 
$\{c_{0}\}_U:=c_{0U}/(\U{m/s})=299\,792\,458$.
Then we obtain
\begin{align}
\vct k
=(1,\,\{c_{0}\}_U)
,
\quad
T=
\begin{bmatrix}
1 & 1
\end{bmatrix}
.
\label{eq:30}
\end{align}
We see that $U\succ V$.
From
$\vct d_1 = (1, -1)\sur T\in{\rm ker}\, T$,
we have
$I_{1U} = \vct k^{-\vct d_1}\vct u^{\vct d_1}=
\{c_{0}\}_U\,\U{m}\,\U{s}^{-1}=c_{0U}$ in $U$,
which is mapped to
$I_{1V}=c_{0V} = 1$ in $V$.
This corresponds to the first step toward natural unit systems \cite{duff}.
This procedure is sometimes written shortly as
``we set $c_0=1$.''

\subsection{MKSA to CGS emu}
Next, we examine a more practical example.
We consider
$U=(\text{MKSA})$ and
$V=(\text{rCGS emu})$.
The base units are
$\vct u=(\U{m},\U{kg},\U{s},\U{A})$ and
$\vct v=(\U{cm},\U{g},\U{s})$, respectively.
Clearly,
we have $\mathcal T(\U{m})=100\,\U{cm}$, and
$\mathcal T(\U{kg})=1000\,\U{g}$.
Using the relation \cite{jackson}:
\begin{align}
\frac{I\sub{emu}}{\U{\sqrt{dyn}}} 
=
\frac{\sqrt{4\pi}}{10}\frac{I\sub{SI}}{\U{A}}
\label{eq:31}
,
\end{align}
or
$\mathcal T(\U{A})=\sqrt{4\pi}10^{-1}\,\U{cm^{1/2}g^{1/2}s^{-1}}$,
we obtain 
\begin{align}
\vct k=(100,\, 1000,\, 1,\,\sqrt{4\pi}/10)
,
\quad
T=
\begin{bmatrix}
1 & 0 & 0 & 1/2\\
0 & 1 & 0 & 1/2\\
0 & 0 & 1 & -1
\end{bmatrix}
.
\label{eq:32}
\end{align}
Clearly, we have $U\succ V$.
With $\vct d_1 = (1, 1, -2, -2)\sur T\in{\rm ker}\, T$,
the representation
\begin{align}
I_{1U}= 100^{-1}\times1000^{-1}\times4\pi\times10^{-2}\,
\U{m\,kg\,s^{-2}\,A^{-2}}
=
4\pi\times10^{-7}\,\U{N/A^2}
=\mu_{0U}
,
\label{eq:33}
\end{align}
in $U$ is identified
with $\mu_{0V} = 1$ in $V$.

\subsection{MKSA to CGS esu}
Similarly, we consider
$U=(\text{MKSA})$ and
$V=(\text{rCGS esu})$.
The base units are
$\vct u=(\U{m},\U{kg},\U{s},\U{A})$,
$\vct v=(\U{cm},\U{g},\U{s})$.
Using the relation \cite{jackson}
\begin{align}
\frac{I\sub{esu}}{\U{\sqrt{dyn}\cdot cm/s}}
= \sqrt{4\pi}\times10\times\{c_{0}\}_U
\frac{I\sub{SI}}{\U{A}}
,
\label{eq:34}
\end{align}
namely,
$\mathcal T(\U{A})=10\sqrt{4\pi}\{c_0\}_U\,\U{cm^{3/2}g^{1/2}s^{-2}}$,
we obtain 
\begin{align}
\vct k=(100,\, 1000,\, 1,\,10\sqrt{4\pi}\{c_{0}\}_U)
,
\quad
T=
\begin{bmatrix}
1 & 0 & 0 & 3/2\\
0 & 1 & 0 & 1/2\\
0 & 0 & 1 & -2
\end{bmatrix}
\label{eq:35}
.
\end{align}
Clearly, we have $U\succ V$.
Using $\vct d_1 = (-3, -1, 4, 2)\sur T\in{\rm ker}\, T$,
the representation
\begin{align}
I_{1U}
&= 100^{3}\times1000\times(4\pi)^{-1}\times\{c_{0}\}_U^{-2}\,
\U{m^{-3}\,kg^{-1}\,s^{4}\,A^{2}}
\NN
&=\frac{1}{4\pi\times10^{-7}\times\{c_{0}\}_U^2}\U{\frac{A^2}{N}\frac{s^2}{m^2}} 
=\frac{1}{\mu_{0U}c_{0U}^2}
=\varepsilon_{0U} 
,
\label{eq:36}
\end{align}
in $U$ can be transferred to
$\varepsilon_{0V} = 1$ in $V$.

\subsection{MKSA to a symmetric three-base unit}
We can construct a three-base unit system
having the symmetry between electricity and magnetism \cite{kennelly, page, kitano}.
We set unit systems $U=(\text{MKSA})$ and $V$ with
$\vct u=(\U{m},\U{kg},\U{s},\U{A})$,
$\vct v=(\U{m},\U{kg},\U{s})$.
We introduce the representation
$Z_{0U}=\sqrt{\mu_{0U}/\varepsilon_{0U}}=c_{0U}\mu_{0U}$
in $U$ of the vacuum impedance $Z_0$ \cite{schelkunoff}.
We can relate the power $P_U$ and the current $I_U$ with the following expression: 
$
P_U = Z_{0U} I_U^2
.
$
Thus, we can express the current with purely mechanical quantities.
Having $\mathcal T(\U{A})=\sqrt{\{Z_0\}_U}\,\U{m\,kg^{1/2}s^{-3/2}}$,
the transformation is given as
\begin{align}
\vct k=(1,\, 1,\, 1,\,\sqrt{\{Z_{0}\}_U})
,
\quad
T=
\begin{bmatrix}
1 & 0 & 0 & 1\\
0 & 1 & 0 & 1/2\\
0 & 0 & 1 & -3/2
\end{bmatrix}
,
\label{eq:37}
\end{align}
where $\{Z_{0}\}_U=Z_{0U}/\U{O}=\{c_{0}\}_U\{\mu_{0}\}_U\sim 377$.
Using 
$\vct d_1 = (2, 1, -3, -2)\sur T\in{\rm ker}\, T$,
the representation
$I_{1U}= \{Z_{0}\}_U\,\U{m^2\,kg\,s^{-3}\,A^{-2}}
=\{Z_{0}\}_U\,\U{O}=Z_{0U}$
in $U$ can be transfered
as $Z_{0V} = 1$ in $V$.

Using this transformation, the set of Maxwell's
equations is unchanged in appearance, although 
the constitutive relations in $U$,
\begin{align}
\vct D_U=\varepsilon_{0U}\vct E_U,
\quad
\vct H_U=\mu_{0U}^{-1}\vct B_U
\label{eq:38}
\end{align}
become 
\begin{align}
\vct D_V = c_{0V}^{-1}\vct E_V,
\quad
\vct H_V = c_{0V}\vct B_V
\label{eq:39}
\end{align}
in $V$.
Followed by the transformation, such as in the second example (Sec.~IX-B), 
into a two-base unit system $\vct w=(\U{m},\,\U{kg})$,
we obtain $c_{0W}=1$ and the following constitutive relations in $W$: 
\begin{align}
\vct D_W = \vct E_W,
\quad
\vct H_W = \vct B_W.
\label{eq:40}
\end{align}

We will refer to $V$ and $W$ as $(\text{MKSA}/Z_0)$ and $(\text{MKSA}/Z_0/c_0)$ 
respectively in this article.
See Fig.~\ref{fig:3}.
This symmetric three-base unit system $(\text{MKSA}/Z_0)$ is rarely used but
it is much simpler than the Gaussian system \cite{page,kennelly}.

\subsection{The modified Heaviside-Lorentz system}

The CGS Gaussian unit system is a mixture
of the emu and esu systems \cite{birge,jackson,kitano}.
In order to satisfy the symmetry between electricity and magnetism,
the two conditions $\mu_{0V}=1$ and $\varepsilon_{0V}=1$ must be imposed simultaneously.
However, it is impossible to satisfy the two conditions
in reducing the number of base units by one, from $N=4$ to $3$.
For such cases, we have usually compromised by choosing one of 
the two unit systems,
the CGS emu and CGS esu,
depending upon the type of quantities involved. 
Given a certain quantity, it is necessary to look up a classification 
list in order to determine which unit system should be applied. 
Unfortunately, there exist several versions of lists.
Here, we use a version referred to as the {\it modified} Gaussian system
\cite{jackson}.
Although not popular, the {\it modified} Gaussian system is more reasonable 
than the widely used version
and can be treated consistently in the current framework.
In addition, we deal with the rationalized version of 
the {\it modified} Gaussian system, which we call
the modified Heaviside-Lorentz (mHL) system.

In order to deal with the mHL,
we must set up a five-base unit system.
Here, we introduce
$U=(\text{MKSAQ})$ with $\vct u=(\U{m, kg, s, A, C})$, where
the dimensions for current and charge
are considered to be independent \cite{sommerfeld}.
The units for electric (magnetic) quantities are derived 
from the unit of charge (current), i.e., the coulomb, ``C'' (ampere, ``A'').

A quantity that relates charge and current;
$\gamma$, the unit of which in $U$ is $[\gamma]_U=\U{C/(A\,s)}$
must be introduced.
Then, the charge conservation can be written as
\begin{align}
\gamma_U^{-1}\fracpd{\varrho_U}{t}=-\diver \vct J_U.
\label{eq:41}
\end{align}
The units for permittivity and permeability are
$[\varepsilon_0]_U=\U{C^2/(N\,m^2)}$ and
$[\mu_0]_U=\U{N/A^2}$, respectively.

The Maxwell equations and the constitutive relations in this unit system are
\begin{align}
& \diver\vct D_U = \varrho_U, \quad 
\curl\vct H_U = \vct J_U + \gamma_U^{-1}\fracpd{\vct D_U}{t}, 
\label{eq:42}
\\
& \diver\vct B_U = 0,\quad
\curl\vct E_U = -\gamma_U^{-1}\fracpd{\vct B_U}{t},
\label{eq:43}
\\
& \vct D_U = \varepsilon_{0U}\vct E_U, 
\quad
\vct H_U = \mu_{0U}^{-1}\vct B_U
.
\label{eq:44}
\end{align}
From these equations, we find the speed of light and the vacuum impedance, as follows:
\begin{align}
c_{0U} = \frac{\gamma_U}{\sqrt{\mu_{0U}\varepsilon_{0U}}}
,\quad
Z_{0U}=\sqrt{\frac{\mu_{0U}}{\varepsilon_{0U}}}
.
\label{eq:45}
\end{align}

We can simply transfer from $U$ to $V=(\text{MKSA})$
using $\mathcal T(\U{C})=1\,\U{A\,s}$.
The transformation is given as
\begin{align}
    \vct k=(1,\, 1,\, 1,\, 1,\, 1)
,
\quad
T=
\begin{bmatrix}
1 & 0 & 0 & 0 & 0\\
0 & 1 & 0 & 0 & 0\\
0 & 0 & 1 & 0 & 1\\
0 & 0 & 0 & 1 & 1
\end{bmatrix}
.
\label{eq:46}
\end{align}
We have $\vct d_1=(0,0,-1,-1,1)\sur{T}\in{\rm ker}\, T$,
and the representation
$I_{1U}=1\,\U{C/(s\,A)}=\gamma_U$ is transferred
to $\gamma_V=1$.

Next, we can transfer from $U$ to $W=(\text{mHL})$, $\vct w=(\U{cm, g, s})$ with
$\mathcal S(\U{A})=10^{-1}\sqrt{4\pi}\,\U{\sqrt{dyn}}$ and
$\mathcal S(\U{C})=10\sqrt{4\pi}\{c_0\}_U\,\U{\sqrt{dyn}\,cm}$.
The transformation is given by
\begin{align}
\vct h=(10^2,\, 10^3,\, 1,\, \sqrt{4\pi}/10,\, 10\sqrt{4\pi}\{c_0\}_U)
,\quad
S=
\begin{bmatrix}
1 & 0 & 0 & 1/2 & 3/2\\
0 & 1 & 0 & 1/2 & 1/2\\
0 & 0 & 1 & -1  & -1\\
\end{bmatrix}
.
\label{eq:47}
\end{align}
We have $\vct c_1=(1,1,-2,-2,0)\sur{T}$,
$\vct c_2=(-3,-1,2,0,2)\sur{T}$, both in ${\rm ker}\, S$,
and corresponding representations
$I_{1U}=\{\mu_0\}_U\,\U{N/A^2}=\mu_{0U}$
and
$I_{2U}=\{\varepsilon_0\}_U\,\U{C^2/(N\,m^2)}=\varepsilon_{0U}$.
These representations are transferred
to $\mu_{0W}=1$ and $\varepsilon_{0W}=1$, respectively.
In addition, we have
\begin{align}
\gamma_W=\mathcal S(\gamma_U)
=\{\gamma\}_U\frac{\mathcal S(\U{C})}{\mathcal S(\U{A})}\,\U{s}^{-1}
=100\{c_0\}_U\,\U{cm/s}
,
\label{eq:48}
\end{align}
namely, $\gamma_W = c_{0W}$
and $Z_{0W}=1$.

Thus, in the modified Heaviside-Lorentz system, 
Eqs.~(\ref{eq:41})--(\ref{eq:44}) are 
changed as follows:
\begin{align}
& c_{0W}^{-1}\fracpd{\varrho_W}{t}=-\diver \vct J_W,
\label{eq:49}\\
& \diver\vct D_W = \varrho_W,\quad
\curl\vct H_W = \vct J_W + c_{0W}^{-1}\fracpd{\vct D_W}{t}, 
\label{eq:50}\\
&\diver\vct B_W = 0,\quad
\curl\vct E_W = -c_{0W}^{-1}\fracpd{\vct B_W}{t},
\label{eq:51}\\
& \vct D_W = \vct E_W, \quad 
\vct H_W = \vct B_U
.
\label{eq:52}
\end{align}
We note that 3-base unit systems,
$(\text{mHL})$,
$(\text{rCGS-emu})$,
$(\text{rCGS-esu})$,
and
$(\text{MKSA}/Z_0)$
are all incomparable (Fig.~3).

In the commonly used Heaviside-Lorentz system and the Gaussian system,
however, $\vct J_W':=c_{0W}\vct J_W$ is used for the current density.
Then, the charge conservation law and the Maxwell-Amp\`{e}re equation become
\begin{align}
& \fracpd{\varrho_W}{t}=-\diver \vct J'_W, 
\label{eq:53}
\\
& \curl\vct H_W = c_{0W}^{-1}\vct J'_W + c_{0W}^{-1}\fracpd{\vct D_W}{t}
,
\label{eq:54}
\end{align}
which seem somewhat irregular with respect to the positions of
$c_{0W}$, compared to those in
Eqs.~(\ref{eq:49})--(\ref{eq:52}) \cite{jackson}.

\subsection{Examples of dimensions}
In this subsection we present several examples showing the
close relation between dimensions and EUS's.

The first example is for the equivalent unit systems.
By replacing the unit of mass, ``$\U{kg}$'' in $U=(\text{MKS})$,
$\vct u=(\U{m},\U{kg},\U{s})$,
with Planck's constant $\hbar=\{\hbar\}_U\,\U{kg\,m^2/s}$,
we can make a new unit system $U'$,
$\vct u'=(\U{m},\hbar,\U{s})$.\cite{post}
With the mapping $\mathcal D = (\vct k, D): U \rightarrow U'$,
we have
$\mathcal D(\U{kg})=\{\hbar\}_U^{-1}\U{m^{-2}}\hbar^1 \U{s}$ and
\begin{align}
\vct k = (1, \{\hbar\}_U^{-1}, 1), \quad
D = 
\begin{bmatrix}
1 & -2 & 0 \\
0 & 1 & 0 \\
0 & 1 & 1
\end{bmatrix}
,
\label{eq:55}
\end{align}
where $D$ is invertible and $U'\sim U$.
Writing the respective dimensions as
$\vct U^{\vct d}$ and
$\vct U'^{\vct d'}$ 
with
$\vct U = (\mathsf L, \mathsf M, \mathsf T)$,
$\vct d = (l, m, t)$,
$\vct U' = (\mathsf L', \mathsf H', \mathsf T')$,
and
$\vct d'= (l', h', t')$,
Eq.~(\ref{eq:13}) becomes
\begin{align}
\begin{bmatrix}
l' \\ h' \\ t'
\end{bmatrix}
=
\begin{bmatrix}
1 & -2 & 0 \\
0 & 1 & 0 \\
0 & 1 & 1    
\end{bmatrix}
\begin{bmatrix}
l \\ m \\ t
\end{bmatrix}
,
\quad
(\mathsf L', \mathsf H', \mathsf T') = (\mathsf L,\mathsf L^2\mathsf M\mathsf T^{-1},
\mathsf T)
.
\label{eq:56}
\end{align}
Both relations can be inverted and
$\vct U^{\vct d}=\vct U'^{\vct d'}$ or
$\mathsf L^l \mathsf M^m \mathsf T^t
=\mathsf L'^{l'} \mathsf H'^{h'} \mathsf T'^{t'}$ always holds.

As an example for the case of $U\succ V$,
we consider $U =(\text{MKSA})$ and
$V = (\text{rCGS emu})$ 
with $\mathcal T:U\rightarrow V$ (See Sec.~IX.C).
We write the dimensions $\vct U^{\vct d}$ and $\vct V^{\vct c}$ with
$\vct U = (\mathsf L, \mathsf M, \mathsf T, \mathsf I)$,
$\vct d = (l, m, t, i)$,
$\vct V = (\mathsf L', \mathsf M', \mathsf T')$, and
$\vct c = (l', m', t')$.
From Eq.~(\ref{eq:32}), we have
\begin{align}
\begin{bmatrix}
l' \\ m' \\ t'        
\end{bmatrix}
=
\begin{bmatrix}
1 & 0 & 0 & 1/2 \\
0 & 1 & 0 & 1/2 \\
0 & 0 & 1 & -1    
\end{bmatrix}
\begin{bmatrix}
l \\ m \\ t \\ i
\end{bmatrix}
,\quad
\mathcal T(\mathsf L, \mathsf M, \mathsf T, \mathsf I) = 
(\mathsf L', \mathsf M', \mathsf T', \mathsf L'^{1/2}\mathsf M'^{1/2}\mathsf T'^{-1})
,
\label{eq:57}
\end{align}
We note that both of which are non-invertible relations.
The dimension in $U$ cannot be derived from that in $V$
and we cannot equate $\vct U^{\vct d}$ and $\vct V^{\vct c}$.

Similarly, we consider the dimension
$\vct W^{\vct b}$ of $W=(\text{rCGS esu})$ (See Sec.~IX.D),
with $\vct W = (\mathsf L'',\mathsf M'', \mathsf T'')$, $\vct b = (l'', m'', t'')$,
and $\mathcal T': U\rightarrow W$.
From Eq.~(\ref{eq:35}), we have
\begin{align}
\begin{bmatrix}
l'' \\ m'' \\ t''        
\end{bmatrix}
=
\begin{bmatrix}
1 & 0 & 0 & 3/2 \\
0 & 1 & 0 & 1/2 \\
0 & 0 & 1 & -2    
\end{bmatrix}
\begin{bmatrix}
l \\ m \\ t \\ i
\end{bmatrix}
,\quad
\mathcal T'(\mathsf L, \mathsf M, \mathsf T, \mathsf I) = 
(\mathsf L'', \mathsf M'', \mathsf T'', 
\mathsf L''^{3/2}\mathsf M''^{1/2}\mathsf T''^{-2})
.
\label{eq:58}
\end{align}
Combining the last two examples, we can see a case of $V \parallel W$.
Each of the dimensions in $V$ and $W$ are derived unidirectionally
from the dimension of $U$.
Neither dimension can be derived uniquely from the other.
In this sense, when two unit systems are incomparable or belong to different EUS's,
their dimensions should be considered unrelated.
For example,
the dimension for electric current in $V$ is
$\mathsf L^{1/2}\mathsf M^{1/2}\mathsf T^{-1}$,
while in $W$ it is
$\mathsf L^{3/2}\mathsf M^{1/2}\mathsf T^{-2}$.
If we equate them, we result in an embarrassing result $\mathsf L\mathsf T^{-1}=1$.
(A related discussion is given in Chap.~VI of Porter's book\cite{porter}.)

\section{Conclusion}
We have investigated the mathematical structures of unit systems and
have found that the set of unit systems can be considered
as a preordered set.
The binary relation $U\succsim V$ implies that
all of the quantities distinguishable in $V$ are
always distinguishable in $U$.
Only in this case there is a mapping from $U$ to $V$, and
the conversion of unit systems from $U$ to $V$ is possible.
We have also found that an equivalence class of 
unit systems (EUS) plays an important role.
There is a {\it partial-order} structure among EUS's 
with the relation $\eus U\succeq\eus V$, 
which is derived from the preorder $U\succsim V$. 
We have also drawn a (partial) hierarchical tree of existing unit systems 
and EUS's in Fig.~\ref{fig:3}.

We have introduced three layers of description of physical
quantities and their representations.
The first layer simply deals with a quantity. 
We denote such a quantity as $Q\in\varOmega$,
which is a rather naive and primitive concept and is completely 
independent of unit systems.
The third layer is the representation $Q_U=q_U\vct u^{\vct d}$
($q_U\in\mathbb R$, $\vct d\in\mathbb Q^N$) of a quantity $Q$ in a unit system $U$.
Although it is a concrete and definite mathematical 
object, the representation is dependent on the unit system.
The intermediate layer is concerned with the e-quantity, 
which denotes collectively all of the representations in 
an EUS as $Q_{\eus U}:=Q_U=Q_{U'}=\cdots$.
The e-quantity is independent of unit systems as long as they belong to the same EUS
and has a definite dimension in the EUS\@.

Generally, formulas and equations in physics should be understood to represent 
the relations of quantities rather than mere numbers. 
In terms of the present discussion,
they specifically represent the relations of e-quantities 
rather than quantities in the naive sense. 
Thus, the result of the present paper provides a theoretical background 
for quantity calculus and dimensional analysis.

In this paper, we have only dealt with scalar quantities.
It is straightforward to
extend to multi-component entities, such as vectors, tensors, 
differential forms and so on.
There, we should not forget to assign dimensions to basis vectors and basis covectors
appropriately, not only to their components.
For example, in the polar coordinate,
using the natural cobasis
$\vct n^r = \nabla r$, 
$\vct n^\theta = \nabla \theta$, 
$\vct n^\phi = \nabla \phi$,
an electric vector field can be represented as
$\vct E=E_r \vct n^r + E_\theta \vct n^\theta + E_\phi \vct n^\phi$.
The dimension of each element is as follows;
$[\vct E] = [E_r]  = \mathsf V\mathsf L^{-1}$,
$[E_\theta] = [E_\phi] = \mathsf V$,
$[\vct n^r] = 1$,
$[\vct n^\theta] = [\vct n^\phi] = [\nabla] = \mathsf L^{-1}$.
where $\mathsf V$ represents the dimension of voltage.
(We assume the EUS containing the MKSA.)
With the cobasis,
the metric tensor is represented as
$\mathsf g = \sum_{ij} g_{ij}\vct n^i\otimes\vct n^j$, where
the dimensions are $[\mathsf g]=[g_{rr}]=1$,
$[g_{\theta\theta}]=[g_{\phi\phi}]=\mathsf L^2$.
As shown in theses examples, the coefficients could have different dimensions.
But dimensions of (co)vectors rectify them and yield the proper dimension
for the vectorial or tensorial quantities, which is independent of
coordinate system.
These are also good examples of the Buckingham Pi theorem.

The conversion from one unit system to another is sometimes troublesome,
especially without a firm foothold. 
The present study reveals clear strategies for unit conversion.
Considering the preorder and partial-order structures and using the normalization,
we can systematically compare the representations in different
unit systems and set up the conversion rules. 
The meaning behind the corner-cutting method in which some quantities
are considered to be unity to move from a unit system to another is clarified. 

In the future, even after old unit systems have been abandoned, 
there will be a need for unit systems other than the present SI, 
which itself may change according as science and technology develop\cite{gupta}.
Therefore, the precise comprehension of the mathematical structure of unit systems
will continue to serve as a theoretical foundation for 
the physical description of nature.

\section*{Acknowledgements}
The authors would like to thank S. Tanimura, Y. Nakata, and S. Tamate for
their helpful discussions.
The present study was supported in part
by Grants-in-Aid for Scientific Research (22109004 and
22560041) and by the Global COE program ``Photonics
and Electronics Science and Engineering'' of Kyoto University.

\end{document}